# A Case Study on Model-Based Development of Robotic Systems using MontiArc with Embedded Automata

Jan Oliver Ringert[*] and Bernhard Rumpe and Andreas Wortmann

Software Engineering
RWTH Aachen University
Ahornstrasse 55
52074 Aachen, Germany
http://www.se-rwth.de/

**Abstract:** Software development for service robotics is inherently complex. Even a single robot requires the composition of several sensors, actuators, and software modules. The systems are usually developed by groups of domain experts, rarely software engineering experts. Thus the resulting software systems are monolithic programs solving a single problem on a single platform. We claim modeling of both structure and behavior of robots in a modular way leads to better reusable software.

We report on a study about the modeling of robotics software with the structure and behavior modeling language MontiArcAutomaton. This study assesses the benefits and difficulties of model-based robotics software development using MontiArcAutomaton. Our findings are based on a survey, discussions with the participants, and key figures from their development behavior. We present the project, our study, lessons learned, and future work based on the insights gained.

## 1 Introduction

We held a one semester course on model-driven robotics software development using MontiArcAutomaton [RRW12], an extension of the architecture description language MontiArc [HRR12] with embedded automata. This course was divided into three stages. During the first stage, the students were assigned to learn about the modeling languages, tooling, and infrastructure based on documentation available. Aim of the second stage was to develop a robotic coffee service. In the third stage, the students improved the MontiArcAutomaton tooling based on the experiences from the second stage.

We conducted this course as a case study on the usage of model-driven engineering (MDE) in the robotics domain in general and on the benefits of using the MontiArcAutomaton modeling language in particular. Using discussions, surveys and key figures from the students' development behavior, we tried to determine whether MontiArcAutomaton can be applied for the development of robotic software and which tools are most essential for successful model based development of robotic software.

[*]J.O. Ringert is supported by the DFG GK/1298 AlgoSyn.

Section 2 explains the course structure, background, and aims. Section 3 reports on the project results of stage two. Afterwards Section 4 describes the results of our survey, discussions with the students, and key figures. Section 5 discusses these results and identifies threats to the validity of our claims. Section 6 highlights related work. Section 7 concludes this contribution with the implications for future work.

## 2   Project Description

During our course, the eight participating master level students learned MDE, development of robot control software, agile development using Scrum, and development of modeling tools. The students were weekly evaluated based on their reports and developed artifacts

The course was divided into three stages: During stage one (9 weeks), the eight participants prepared presentations on technologies and practices, e.g., Scrum, JUnit, MontiCore [KRV10, Sch12], and MontiArcAutomaton, to be applied during development. In stage two (6 weeks) the students developed a system of robots able to provide a coffee service using Lego Mindstorms robots. The robots development stage ended with a presentation of the working system[1]. Afterwards, we surveyed the students on their modeling experiences during the development stage. In the third state (10 weeks), the students were assigned to improve the MontiArcAutomaton tooling based on their answers from the survey and discussions.

During the latter two stages, the team was led by a student ScrumMaster and a student Product Owner. We posed as Customer and User providing requirements and feature decisions. We participated in weekly sprint meetings to plan the next sprint and review the previous one. Due to the participants schedules, the development team was unable to have daily Scrum meetings. Weekly participation was mandatory to pass the class.

**Goals**   The goals of our study are to (1) determine whether MontiArcAutomaton can be applied for the development of robotic control software, (2) which tools are most essential for successful model-based development of robotic software. More detailed, we want to determine (1a) whether the decomposition of a robotic system can be adequately modeled using MontiArc and whether (1b) the control logic and behavior of components can be adequately modeled using the I/O$^\omega$ automata paradigm [Rum96, RRW12] with message passing based communication. We also wanted to find out (2a) whether the existing tools for code generation and context condition checks were missing any features and (2b) which tools were missing for effective and efficient development of robotic software.

**Students and Background**   All of the participating students had selected our course as first choice out of three choices. All but one student had previously attended the lecture Software Engineering with a basic introduction of model-based software development. Five of the students had attended the lecture Model-Based Software Engineering on UML and other modeling languages. Two of the students had attended the lecture Generative Software Engineering on DSL and code generator development. Three of the students had experience with the Lego NXT platform. Two of the students are professional mathematical technical software engineers and have programming experience from industry jobs. One of the students created his bachelor thesis already using the MontiCore code generation framework and the architecture description language MontiArc.

---
[1]See the video at http://www.se-rwth.de/materials/ioomega/#RoboticsLab

None of the students had experience with the modeling language MontiArcAutomaton and the runtime framework developed for the Lego NXT robots. In the first stage of the project each student was assigned a topic to prepare a presentation in front of the group. The topics covered in the presentations were (please refer to our website [WWW13] for more information and the presentation slides): development tools (SVN, JUnit), leJOS, MontiCore (AST generation), MontiCore (language composition), MontiCore (code generation and templates), MontiArc, MontiArcAutomaton language, and MontiArcAutomaton code generation.

**Available Material** For the first stage of the course we supplied our students with technical reports, papers, and presentations about the technology and tools to use. The materials included an introduction to the modeling language MontiArcAutomaton. MontiArcAutomaton may be applied to describe the logical decomposition and physical distribution of Cyber-Physical Systems as a component and connector model. The behavior of components can either be modeled as I/O$^\omega$ automata or implemented in a general-purpose programming language (currently MontiArcAutomaton supports Java and Python [RRW13]). In addition, behavior of components can be defined as the composition of existing components. All components exhibit their typed and named ports and a Javadoc-like documentation, but not their implementation details. This uniform handling of components makes the composition mechanisms independent of component implementations.

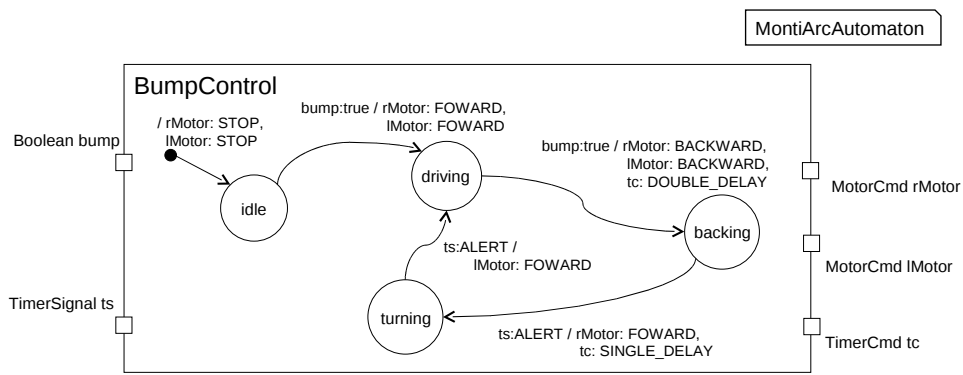

Figure 1: A MontiArcAutomaton model from [RRW13] of component `BumpControl` with an embedded automaton. Transitions read messages, e.g., sensor data, from typed input ports and send messages on typed output ports.

Figure 1 shows the component `BumpControl` with the incoming ports `bump` of type `Boolean` and `ts` of type `TimerSignal` (on the left) and the outgoing ports `rMotor`, `lMotor`, and `tc` (on the right). The input and output behavior of the component `BumpControl` is modeled by the I/O$^\omega$ automaton shown inside the component. In every execution cycle of the system transitions are activated by a matching pattern of incoming messages and their execution emits messages on ports as specified in the output block. The types of messages are either defined in Java or UML/P [Rum11, Sch12] class diagrams. MontiArcAutomaton supports local variables with assignments on the output blocks of transitions and OCL/P [Rum11] guards on transitions. More details on MontiArcAutomaton are available from [RRW12, RRW13, WWW13].

For the second stage the students were supplied with two Lego NXT education kits and additional four regular Lego NXT sets. The NXTs were running Java using the leJOS[2] firmware. We provided a code generator for MontiArcAutomaton models to Java code and a runtime environment with a library of platform specific components for the NXT leJOS platform. The students were using services of the SSELab [HKR12] that include a wiki, a bug tracking system, a mailing list, and a SVN repository to collaborate.

**Task** The task of the students in the development stage (second stage) was to develop a system of autonomous robots that is able to receive coffee orders and deliver coffee to different offices and places. The initial task was described in 10 user stories. An example of a user story is *"As coffee with milk drinkers we want the robot to be able to fetch coffee with milk."*. All user stories are available from [WWW13]. We gave no explicit orders on the number of robots to develop and many of the design and implementation decisions were left open, e.g., the implementation of ordering a coffee via a robot, web browser, or smart phone. Following Scrum practice the requirements were elicited, refined, and discussed in regular meetings.

**Analysis Procedure** Our analysis of the project is based on informal discussions with our students, results of a questionnaire, and statistics of their development behavior based on a version control system. In weekly meetings we had discussions about problems encountered during each previous sprint. We documented these discussions, which contained technical or organizational problems of the team, as informal notes. After the development stage and the presentation of the running system we discussed the deficiencies of the existing tools and the necessary improvements and additional tools to be created.

To capture the subjective evaluation of our students we created a questionnaire filled out after the development stage. The questionnaire contains 14 questions about the efforts spent on learning and development, about the confidence in the models and implementations, about the effort of fixing bugs and the amount of testing.

As surveys are subject to several issues, we also studied the students' development behavior. Therefore we identified several interesting key figures and monitored these from Oct. $21^{st}$ to Nov. $17^{th}$ at a three days sampling rate. The key figures identified are number of architecture changes in composed components, behavior changes in I/O$^\omega$ automata, and changes in Java files as well as the total number of composed and automaton components and number of components with Java implementations.

We furthermore counted for each project the number of components that were defined and instantiated within this project and the number of components from a library or runtime environment that were instantiated within this project in the final robot implementations.

---

[2]LeJOS Project website: http://lejos.sourceforge.net/

## 3 Project Results

The students solved the second stage by implementing a system of three cooperating robots, that receive a coffee order via a website, pick up a mug, fetch coffee, and deliver it to the person having placed the order. The system consists of a mug provider robot, a coffee preparation robot, and a coffee delivery robot as depicted in Figure 2. The coffee preparation robot is connected to an Android cellphone via Bluetooth, which hosts the coffee service website. After a user requests a coffee via the website, this issue is forwarded to the coffee preparation robot, which informs the coffee delivery robot to pick up a cup, fetch coffee, and deliver it to the user.

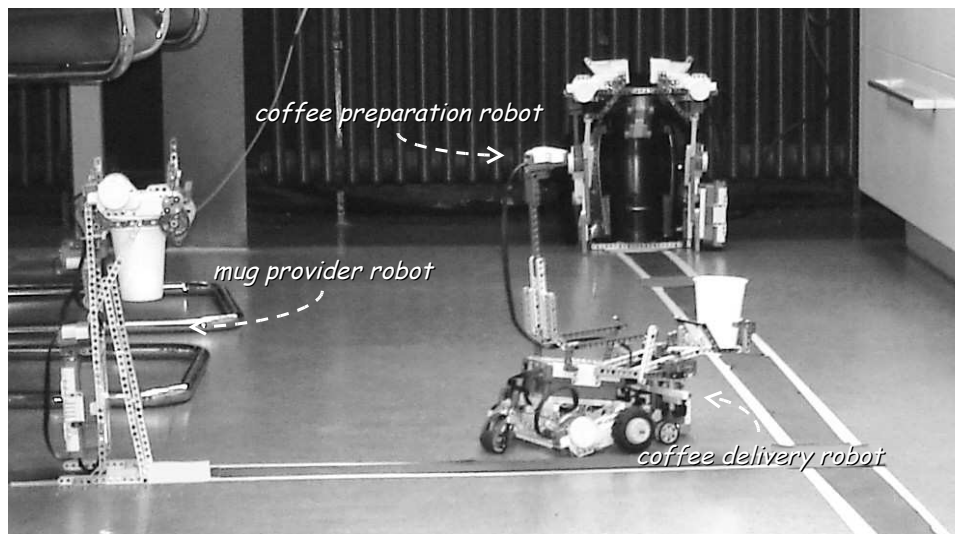

Figure 2: The coffee service at work. A coffee request triggered the coffee service robot to pick up a mug and proceed to fetch coffee. A video is available from our website [WWW13].

Structure and behavior of all robots were modeled using MontiArcAutomaton. From these models, Java implementations were generated using MontiCore. The application models interface the robot using library components wrapping respective parts of the leJOS API.

The coffee delivery robot consists of 10 original component models. Four of these contain I/O$^\omega$ automata and four others are composed. The remaining two component models have Java implementations. The robot further reuses 15 component models from the library, e.g., wrappers for hardware sensors and actuators. Figure 3 shows the composed component `NavigationUnit` implementing the navigation of the robot. The component `NavigationController` for example determines the robot's next actions using an I/O$^\omega$ automaton, the input from several buttons (`E,L,R`) and augmented readings from two color sensors (`SR,SL`). The component `DifferentialPilot` is implemented in Java and uses a part of the leJOS framework to move the robot. The atomic component `NavigationControl` contains an I/O$^\omega$ automaton of the same name, which consists of 11 states and 41 transitions featuring OCL/P guards.

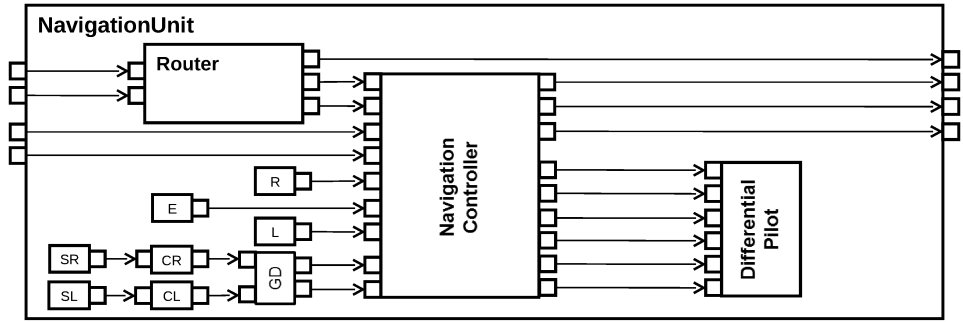

Figure 3: Structure of the composed navigation management component `NavigationUnit`.

## 4 Analysis and Interpretation

Our analysis covers results from discussions, model changes from version control, and a survey as described in Section 2.

### 4.1 Results from the Weekly Discussions

In weekly meetings, we conducted regular discussions with the students about the issues at hand. During the development stage, the students had problems structuring the development tasks which yielded claims for better communication and for a dedicated system architect administering interface changes. Due to the lack of arrangement, the Bluetooth communication was implemented three times: (1) as a detailed design document, (2) as a deviating implementation, communicating complex data structures not supported by MontiArcAutomaton, and (3) as another deviating implementation, communicating enumeration types. Similarly, the initial architecture was dismissed unused. Another important issue was the lack of tool support to facilitate development. The students claimed that editor support (e.g., model completion, context conditions) is crucial to efficient MDE and thus created a text editor with these features in the third project stage. Other technological issues expressed were the parallel development of both software and hardware, the complex Maven[3] build cycle, and the manual deployment of components to platforms.

Regarding MontiArcAutomaton our students had initial problems to restrict component communication to simple message passing using automata. On the other hand, they did not miss language features like hierarchical states in I/O$^\omega$ automata. This supposedly is due to the existence of composed components, which allow similar decomposition mechanisms using subcomponents. Please note, that 7 students attended the Software Engineering class teaching basic concepts of Statecharts while 5 students in addition attended the Model-Based Software Engineering class with exercises on more complex features, e.g., history states or entry actions. We are not aware of Statechart modeling experience of the students beyond these lectures.

---

[3]Maven project website: http://maven.apache.org/.

On a less technical note, the students mentioned that Scrum may be sub-optimal when the semester schedule prohibits daily Scrum meetings. Unfortunately, this issue is beyond our control. While the discussions mostly addressed process issues, the questionnaire focused on the effort required to model robot control software using MontiArcAutomaton.

## 4.2 Results from the Questionnaire

After the students finished the development stage with a presentation of the running system, we conducted a survey using the questionnaire available on the MontiArcAutomaton website [WWW13]. We asked the students 14 questions on their development efforts, problems, confidence in the artifacts, and testing behavior. We handed out eight questionnaires and all eight were returned fully filled.

The results suggest that learning MontiArcAutomaton was a major effort during the development stage of the course and that this effort can be eased by helpful development tooling (e.g., editor support with early feedback). The students reported to have implemented only two regression tests, but many manual tests of their systems.

First, we asked the students how much time they have spend learning our technologies. The results, displayed in Figure 4 (a), point out that learning the leJOS robot API was significantly easier than learning MontiArcAutomaton. We believe this is partly due to the amount of available well-written documentation for leJOS, many examples, and an active online community. Compared to this, information on MontiArc and MontiArcAutomaton was only available from two technical reports, one paper, and about 30 example models.

The second question of the questionnaire asked what percentage of the time the students spent on using MontiArc, MontiArcAutomaton, leJOS, or the MontiCore code generation. Figure 4 (b) shows, that the students spent most of their time modeling the behavior of components, followed by the time spend to construct the actual Lego robots. While the former did not surprise much, we learned from observations that it was surprisingly hard to construct useful Lego robots without building instructions.

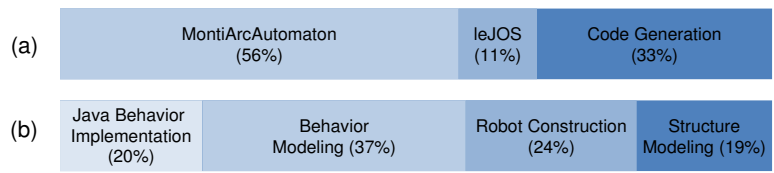

Figure 4: Fractions of the time spent on learning the technologies (a) and on creation of the three robots of the coffee system (b).

We asked the students how they estimated the efforts required to understand the component models, automaton models, and Java implementations on a scale from 1 (simple) to 10 (almost impossible). The results displayed in Figure 5 (a) show that the students were able to comprehend MontiArc as well as Java, which they have been introduced to in their first bachelor semester. The I/O$^\omega$ automata were considered more complex and harder to understand. While we expected similar results, we were surprised by the students' feedback that the effort for fixing bugs[4] in the component models was assumed to be as high as the effort for fixing bugs in I/O$^\omega$ automata (Figure 5 (b)). The students considered the effort for fixing bugs in MontiArcAutomaton twice as high as the effort for fixing bugs in Java artifacts. Discussions yielded the result that this is due Java being taught from the first semester and due to the lack of development tooling for MontiArcAutomaton.

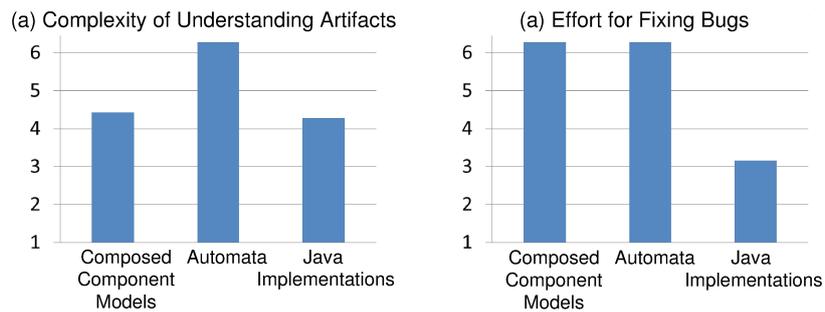

Figure 5: Efforts for understanding the different development artifacts and fixing bugs as rated by the students on a scale from 1 (simple) to 10 (almost impossible).

We also asked the students to estimate their confidence in the artifacts created by them and their team members on a scale from 1 (no confidence) to 10 (works perfectly). While we found little difference between the confidence in the artifacts created by themselves or others, the students were overall less confident in the I/O$^\omega$ automata than in the component and Java artifacts, which follows from the assumed complexity of I/O$^\omega$ automata.

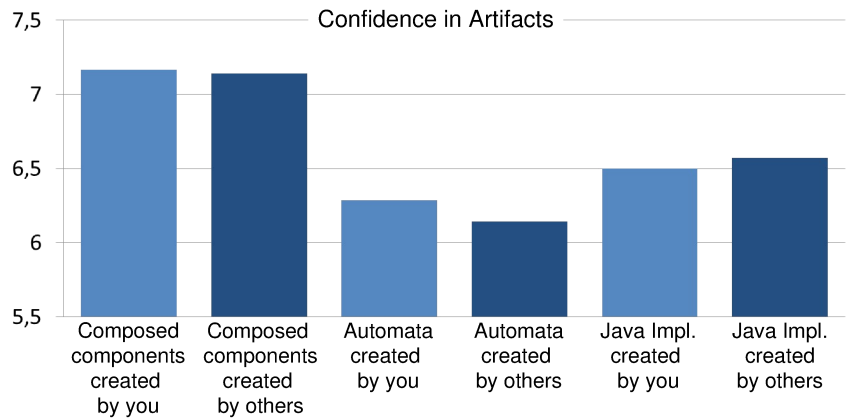

Figure 6: Confidence in the correctness of different development artifacts as rated by the students on a scale from 1 (no confidence) to 10 (works perfectly).

The students claimed to have used I/O$^\omega$ automata for 57.3% of the atomic components and that they could have used I/O$^\omega$ automata for up to 64.6% of the components. Actually they modeled 12 of the 23 (= 52.2%) components with I/O$^\omega$ automata and developed two components with Java implementations. If these would have been modeled with I/O$^\omega$ automata too, they would have modeled the total of 60.1% components with I/O$^\omega$ automata, which is close to their estimate. In conclusion, the students indirectly estimated, that they could have modeled all behavior implementations with I/O$^\omega$ automata. Overall, the results showed that learning a new modeling language poses the expected challenges and is even harder if lacking tool support.

---

[4]Our broad definition of *bug* covers any incorrect or unexpected behavior of the software (sub-)system.

### 4.3 Results from the Students' Development Behavior

As questionnaires are subject to several issues, we also monitored the students' development behavior as described in Section 2. We present several interesting key figures.

**Component implementations: Models vs. Java** For the three robot projects we have counted the numbers of MontiArcAutomaton components implemented as pure models (automata and composed components) and the number of components implemented in Java over time as shown in Figure 7. The rise after six days was an initial version of the robots created by the students mainly in Java. Another week later the components implemented as composed and automaton components started to outgrow the Java implementations. Of the two remaining Java implementations one is a value lookup component and the other one implements an obstacle detection and classification. Review suggests that both could be implemented using I/O$^\omega$ automata with some additional effort. The development over time indicates the learning curve of our students that initially created the implementation in Java and then switched to modeling all but two components.

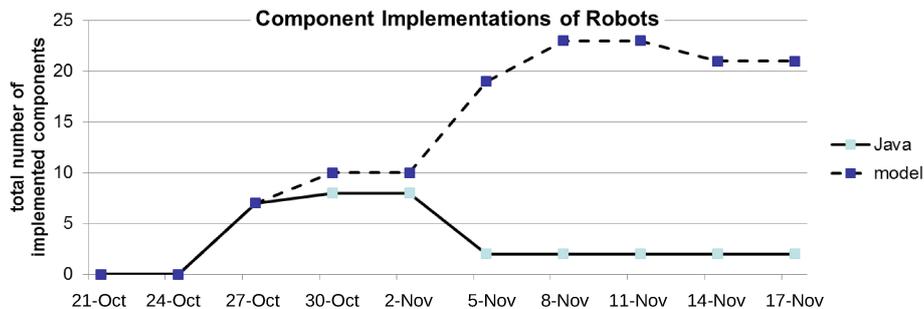

Figure 7: The total number of components implemented as Java implementations and models.

**Changes of artifacts: Models vs. Java classes** We have analyzed the changes of artifacts in the version control history of the students' projects. We distinguish between the library project with hardware wrappers implemented in Java and the three projects containing the robot control software.

The numbers of changes per artifact in the three robot projects are shown in Figure 8. Again we can see a difference from the beginning of development – with equal or more changes per Java file – to the second part of development with many more changes per model file. We see a peak of 2.4 changes per MontiArcAutomaton model around Nov. $8^{th}$ before the scheduled *final* presentation of the robots. On average and over the complete development time each MontiArcAutomaton model has been changed 8.6 times and each Java file 4.8 times.

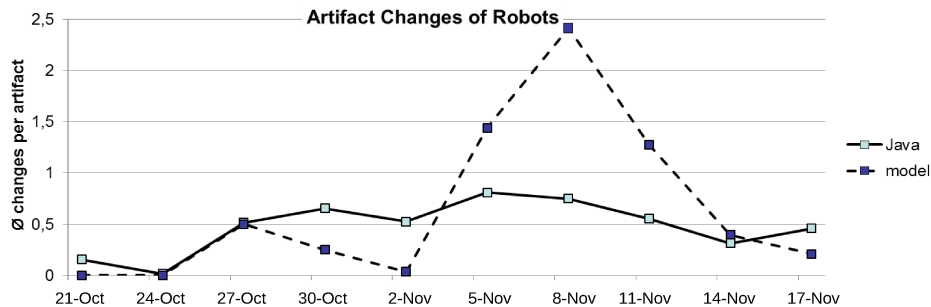

Figure 8: The changes over time per Java and model file of the implementations of all three robots.

An interesting observation on the artifact changes in the library project is that models were changed on average 0.3 times while each Java file was changed 4.5 times. This indicates that component interfaces did not change much while component implementations did.

# 5 Discussion

Modeling robot control software poses several challenges. In both discussions and survey, the students pointed out that learning the I/O$^\omega$ automaton part of MontiArcAutomaton was rather hard. On one hand one could expect that I/O$^\omega$ automata with their similarity to Mealy machines or UML Statecharts are easy to learn, on the other hand mastering a new language and paradigm for application development is never easy. This problem was amplified due to the lack of tooling supporting the modeling process. Despite these issues, the students developed 86% of the atomic components using MontiArcAutomaton and believe that they could have modeled the remaining atomic components also.

The generalizability of this estimation is debatable as solutions to other common robotics problems (e.g., trajectory planning) may not be modeled using I/O$^\omega$ automata. MontiArcAutomaton is implemented as a MontiCore language, which allows to embed arbitrary MontiCore behavior modeling languages into components. Thus, specific modeling languages for common robotics problems may further ease robotics software development.

During the development stage, the students especially claimed that editor support is crucial to efficient MDE. They therefore developed a text editor with several features and integrated it into a graphical editor Eclipse plugin for I/O$^\omega$ automata[5]. To facilitate deployment, they further developed a deployment language profile of MontiArcAutomaton, which maps components to platforms and communication technologies.

---
[5] See the video at http://www.se-rwth.de/materials/ioomega/#Editor

### 5.1 Threats to Validity

We performed this study on a single software project with a group of eight master level students. The students took part in this course to obtain a certificate and data was gathered using discussions, a survey, and development figures. This setup yields threats to the internal validity (causality) and to the external validity (generalizability) of this study.

Threats to the internal validity stem from the students' lack of previous experience using MontiArcAutomaton. They had to learn and apply the MontiArcAutomaton language before they could determine any benefits for modeling robotics software. Another study, wherein the same group of students develops a new robot control software with prior knowledge of MontiArcAutomaton could resolve this issue. Further threats to causality follow from the instruments we chose to determine the students opinion on modeling: Questionnaires are subject to several issues, e.g., the scales are understood differently by participants, there are several well-known response biases, and the results only reflect the participants self-perception. While the former two issues are inherent problems of from the students' development behavior as explained in Section 4.

Threats to generalizability ensue from the number of participating students and the fact that the students were graded. While the first threat can be eliminated by future experiments with a greater number of students, omitting grades is not feasible in the setting of a university course.

## 6 Related Work

Research in robotics software engineering has led to several frameworks for component based development [BKM+05, BBC+07, SSL11], modular architectures [QGC+09, LW11], and robot behavior modeling [Mur02, WICE03]. MontiArcAutomaton differs from these approaches by integrating modeling of structure and behavior. MontiArcAutomaton is furthermore implemented as a MontiCore language, thus facilitating the embedding other behavior modeling languages and integration of other code generators [RRW13].

AutoFOCUS [BHS99, HF07] is a complete IDE with many graphical modeling languages for specification, implementation, testing, and deployment of reactive distributed systems. AutoFOCUS' semantics for automata is similar to the one currently implemented in our MontiArcAutomaton robotics framework. We are not aware of code generators from AutoFOCUS models to robotic platforms or reports on studies similar to ours.

As our students have noticed, efficient robotics software development requires a supporting toolchain. While there exist numerous development environments [KS07, Mor08, WNBG12] none of these has yet gained a similar popularity as, e.g., Eclipse in the software engineering domain.

There are several case studies on the application of modeling techniques to general software development [BSGS04, Sta06, MD08, KJB+09] or to specific domains [KKP+06, BE08]. These describe how the authors applied MDE to a certain problem or domain and thus are biased in a sense, that the authors use their tools and solutions themselves. In contrast, our results are based on the feedback of students new to MDE. Thus, this feedback takes the entrance barrier for software developers new to MDE into account.

The only similar study we found discusses the effects of teaching modeling techniques to students during a university software project course [CAWK09]. The authors conducted a course on MDE with a group of 23 students, which finished with a single survey to determine whether software development benefits from MDE. As the authors did not control whether the students' answers match their behavior, the questionnaires results solely rest on the students' self-perceptions. This study further did not focus on robot control software. Although their survey focused on the understanding gained for different aspects of MDE from actually using it, their results point into the same direction as ours.

## 7 Conclusion

We have learned that documentation and editor support are the most important artifacts to support the introduction of a new modeling language. While peripheral tools (e.g., concerning building and deploying the software) might increase the development efficiency, they should not increase the toolchain complexity.

The students intensively used composed and atomic MontiArcAutomaton components to model structure and behavior of the robots. During the development stage, they even changed almost all atomic components from Java implementations to MontiArcAutomaton models. Thus we believe that MontiArcAutomaton is suitable for the development of robotic control software. The coffee delivery system consists of 60 component instances of which 39 are generic library components. This shows the modularity and reusability supported by MontiArcAutomaton.

The validity of our findings should be tested in future studies to reduce the threats identified in Section 5. We propose to set up consecutive studies with greater numbers of participants and in two groups, where both groups should develop the same robot control software. The control group should be assigned to develop the system using a general purpose programming language unknown to them, while the experimental group should be assigned to model the system using a modeling language also unknown to them.

**Acknowledgments**

We thank our students who participated in the study and congratulate them on winning the audience award for the best student project presented at the Software Engineering 2013 conference[6].

---

[6]See http://www.se-rwth.de/news/ entry of 2013-02-27.